\documentclass[twocolumn,prc,showpacs,amsmath,amssymb]{revtex4-1}

\usepackage{graphicx}

\begin{document}

\title{Structure of the phonon vacuum state}

\author{S. Mishev}
\email{mishev@theor.jinr.ru}
\affiliation{Joint Institute for Nuclear Research, 6 Joliot-Curie str., Dubna 141980, Russia}
\altaffiliation{On leave from Institute for Nuclear Research and Nuclear Energy, Bulgarian Academy of Sciences}
\date{\today}

\begin{abstract}
  The action of the long-range residual force on the expectation value
  of observables in the nuclear ground states is evaluated by finding
  optimal values for the coefficients of the canonical transformation
  which connects the phonon vacuum state with the (quasi-)particle
  ground-state. After estimating the improvements over the predictions
  of the independent-particle approximation we compare the
  ground-state wave functions, obtained using the presented approach,
  with those, obtained using the conventional random phase
  approximation (RPA) and its extended version. The problem with
  overbinding of the nuclear ground state calculated using the RPA is
  shown to be removed if one sticks to the prescriptions of the
  present approach. The reason being that the latter conforms to the
  original variational formulation. Calculations are performed within
  the two-level Lipkin-Meshkov-Glick model in which we present results
  for the ground and first excited state energies as well as for the
  ground-state particle occupation numbers.
\end{abstract}



\maketitle

\section{Introduction}

Approximating the correlated ground state of a many-body quantum
system has been receiving considerable attention since the early days
of nuclear structure physics \cite{1960_baranger} and degenerate
electron gas theory \cite{1957_sawada} and still represents a
formidable challenge. This is an arduous task within the ``beyond the
mean-field'' theories because of the action of the residual
interaction which brings particle-hole admixtures into the ground
states. In the present paper we focus on the effects of long-range
part of this interaction. Unambiguously attributable to the latter is
the quadrupole correlation energy, which, as shown by the findings in
Ref.\cite{2006_bender}, is considerable and varies between 100 keV and
5.5 MeV in different nuclei which makes the perturbative treatment
unsuitable. The short-range residual forces, on the other hand,
compete with the long-range ones in dominating the ground-state shapes
formation \cite{1970_rowe, 2012_qi}. As a result of this competition,
in the beginning and the end of major shells the nucleons are paired,
giving rise to spherical shape, while in the middle of the shell the
nucleons are paired-off and they align to the field generating forces
thus contributing to deformation. We conjecture that the present study
can serve as a foundation to investigate the mechanism of the
transition between these two regimes and in particular on the pairs
decoupling process. The process that we surmise was concocted in
Ref. \cite{1964_hara} and essentially implies that the long-range
force breaks nucleon pairs which may further recouple due to the
pairing force.

We approximate the nuclear ground-state wave function with the phonon
operators \cite{1992_soloviev} vacuum state. A general form of the
phonon vacuum was proposed by Sorensen \cite{1961_sorensen} and later
Goswami and Pal \cite{1963_goswami} estimated explicitly the
correlation coefficients of the $2p-2h$ admixtures into the BCS wave
function \cite{1957_bcs} relating them to the forward and backward
phonon amplitudes. The relation they obtained turned out to be also
valid for higher order correlations \cite{1965_sanderson} in the
random phase approximation (RPA) \cite{1960_baranger}. Being a small
amplitude limit of the time-dependent Hartree-Fock approximation,
however, RPA is known to be able to account only for small correlation
effects. Since in open-shell nuclei the backward phonon amplitudes are
by no means small, RPA is becoming questionable in describing the
low-energy states of such nuclei. On a quest to construct a
microscopic global theory for the nuclear binding energies it was
shown in Refs. \cite{2000_hagino, 2002_stetcu} that the RPA is a
useful method around spherical and well deformed nuclei but falls
short in describing nuclei from the transitional region. This problem
was addressed by Hara \cite{1964_hara} and later on by Jolos and
Rybarska \cite{1980_jolos}, who proposed an improvement over the RPA,
referred to as extended RPA(ERPA), based on the Pauli blocking
principle which plays a progressively important role with the
increasing number of the valence nucleons. This extension has proven
to be in better accord with the experimental data as demonstrated in a
number of papers as for example in Refs.\cite{1993_karadjov,
  2010_mishev}.  Although its superiority over the standard RPA is
undeniable, the variational character of the theory is violated as the
ground state is found to be overbound. The strong argument that the
variational property of a theory insures a converging succession of
approximations to the exact solution fostered the formulation of a
elaborate formalism, called self-consistent RPA \cite{1990_dukelsky},
which as in the conventional particle-hole theory, allowed to take
into account the nucleon correlations without explicitly constructing
the ground-state wave function. In Ref.\cite{1999_hirsch} within the
schematic two-level Lipkin-Meshkov-Glick(LMG) model \cite{1965_lipkin}
it is found that the self-consistent theory allows one to go beyond
the point of RPA collapse, but near this point the wave function of
the first excited state within this approximation is found to be
almost orthogonal to the exact one. This behavior is due to the higher
order particle-hole admixtures contributing to the structure of the
first excited state.

We note that beside the aforementioned amendments to the standard RPA,
there exist a number of other important developments on this subject
as, for example, those in Refs. \cite{1968_rowe, 1990_lenske,
  1994_catara, 1996_catara, 2000_grasso, 2004_kuliev, 2005_delion}.

In the present paper we keep using the explicit form of the fermionic
many-body vacuum \cite{1961_sorensen} but depart from varying the
excited-state wave function. On the contrary we use the correlation
coefficients as parameters which we fix by optimizing the ground-state
trial wave function using a variational procedure. The excited states
corresponding to such vacuum state are of $1p-1h$ type. This approach
also benefits from the findings in Ref.\cite{2010_jemai} where it was
shown that this class of wave functions is a vacuum for a generalized
phonon operator, adding to the standard one specific two-body
operators correcting for the Pauli principle. By way of example, using
the two-level LMG model, they showed that the additional terms improve
the convergence substantially. In this way the phonon vacuum state
absorbs additional correlations effects than the ones obtained using
the equations-of-motion method \cite{1970_rowe} for the standard
phonon operator.

The paper proceeds as follows. In Section \ref{sec:formalism} we
outline the problem and summarize the main obstacles towards the exact
solution. Basic equations of several approximate methods including the
RPA, ERPA and the explicit variation of the phonon vacuum state along
with the exact solution within the LMG model are derived in Section
\ref{sec:lmg}. A comparison between them is established on the basis
of the ground and the first excited state energies as well as
  the on ground-state particle occupation numbers in Section
\ref{sec:numerics}. Summary and outlook is given in Section
\ref{sec:conclusion}.

\section{Formulation}
\label{sec:formalism}

Formally a wave function which contains admixtures into the
independent-quasiparticle wave function and is a vacuum for the phonon
operators \cite{1992_soloviev}
 
\begin{equation}
  \label{eq:def_phonon}
  Q_{\lambda \mu i}=\frac 12\sum_{11^\prime
  }\left[\psi_{11^\prime}^{\lambda i}A(11^\prime|\lambda
    \mu)-(-1)^{\lambda -\mu }\varphi _{11^\prime}^{\lambda
      i}A^\dagger(11^\prime |\lambda -\mu )\right] 
\end{equation}

can be expressed as \cite{1961_sorensen}

\begin{equation}
  \label{eq:def_transform_gs}
  |\rangle = N_{0}. e^{\hat{S}} \rangle
\end{equation}

with

\begin{equation}
  \label{eq:oper_nqp}
  \hat{S}=- \frac 12  \sum_{12; \lambda \mu} S_\lambda(11^{\prime};22^{\prime}) (-)^{\lambda-\mu} A^\dagger(11^{\prime}|\lambda \mu) A^\dagger(22^{\prime}|\lambda -\mu).
\end{equation}

Here and below an independent shell-model state $\lbrace
N_{1},l_{1},j_{1},m_{1} \rbrace$ is abbreviated as $1$.

The coefficients $S_\lambda(11^{\prime};22^{\prime})$
are referred to as correlation coefficients and denote the amplitudes
for the presence of zero, four, eight, ... quasiparticles in the
ground state due to the virtual vibrations.  These present a primary
source of structure information for the ground states and make up a
major part of our present research. The correlated and uncorrelated
ground states are denoted as $|\rangle$ and $\rangle$
correspondingly. The two-quasiparticle creation operator
$A^\dagger(12 ; \lambda \mu)$ reads
\begin{equation}
  \label{eq:A}
  A^\dagger(12 ; \lambda \mu) = \sum_{m_1 m_2}\langle j_1 m_1 j_2 m_2 | \lambda \mu \rangle \alpha_{j_1 m_1}^\dagger\alpha_{j_2 m_2}^\dagger,
\end{equation}

where $\alpha_{jm}^\dagger$ denotes the quasiparticle creation operator.

In the following we shall also need the quasiparticle scattering
operator

\begin{equation}
  \label{eq:B}
  B(12 ; \lambda \mu) = \sum_{m_1 m_2}(-)^{j_2+m_2}\langle j_1 m_1 j_2 m_2 | \lambda \mu \rangle \alpha_{j_1 m_1}^\dagger\alpha_{j_2 -m_2}.
\end{equation}

Details on the quasiparticle-phonon nomenclature, which we follow in
this paper, can be found in Ref.\cite{1992_soloviev}.

Using the ERPA, the correlation coefficients are found to satisfy the
equations \cite{1961_sorensen, 1964_hara}

\begin{equation}
  \label{eq:S_rpa_psi_phi}
  \varphi_{11^{\prime}}^{\lambda i} = \sum_{22^{\prime}}(1-\rho_{22^{\prime}})S_{\lambda}(11^{\prime}|22^{\prime})\psi_{22^{\prime}}^{\lambda i},
\end{equation}

where $\rho_1$ is the quasiparticle occupation density on the level $1$.

In Eq.\eqref{eq:def_transform_gs} $N_{0}$ is a normalization factor which in
physical terms is the overlap between the independent-particle
and the correlated wave functions. It is found to be 

\begin{equation}
  \label{eq:normalization_factor}
  N^{2}_{0} = \frac{1}{\langle e^{(S^{\dagger}+S)} \rangle}.
\end{equation}

In RPA, suggesting small correlations so that higher order terms
contribute relatively little, this constant is approximated as
\cite{1962_brown,1965_sanderson}:

\begin{equation}
  \label{eq:normalization_factor_appr}
  N^{2}_{0} \approx \frac{1}{e^{\frac 12 \sum_{11^{\prime}22^{\prime};\lambda}\pi_{\lambda}^{2}S^2_\lambda(11^{\prime}|22^{\prime}) } }.
\end{equation}

Here and below we use the shortcut notation
$\pi_{\lambda}=\sqrt{2\lambda+1}$, with
$\pi_{\lambda\lambda^{\prime}}=\pi_{\lambda}\pi_{\lambda^{\prime}}$. 
 An explicit solution to the system
\eqref{eq:S_rpa_psi_phi} was obtained by Hara \cite{1964_hara}.

Changing the frame of mind we shall try to obtain the correlation
coefficients by explicitly varying the wave function $|\rangle$ in the
functional 

\begin{equation}
  \label{eq:varational_principle}
  \delta \langle | H | \rangle = 0
\end{equation}

, with $S_\lambda(11^{\prime};22^{\prime})$ being
  variational parameters, i.e.  we shall try to solve the
  equation

\begin{equation}
 \label{eq:eq_general}
  \delta(N_{0}^{2} \langle e^{S^{\dagger}} H e^{S} \rangle) \equiv \langle e^{S^{\dagger}} H e^{S} \rangle \left( \delta N_{0}^{2}\right) + N_{0}^{2}\left( \delta  \langle e^{S^{\dagger}} H e^{S} \rangle \right) = 0,
\end{equation}

with respect to $S_\lambda(11^{\prime};22^{\prime})$.  Assuming that
states with more than four quasiparticles are less probable to be
excited due to the pairing gap \cite{1960_baranger}, i.e. the
configuration space for the ground state is restricted to four
quasiparticle admixtures only, the quantities that need to be
evaluated are presented in the following expression

\begin{equation}
  \label{eq:energy_expansion}
  N_{0}^{2} \langle e^{S^{\dagger}} H e^{S} \rangle \approx \frac{ \langle H \rangle + 2 \langle H S\rangle + \langle S^{\dagger} H S\rangle  }{1 + \frac 12 \langle S^{\dagger}S \rangle}.
\end{equation}

The validity of this assumption is examined in
Sec. \ref{sec:numerics}, using a simplified setup provided by the LMG
model, where it is shown that it holds true at strengths weaker than or in
the vicinity of the RPA point of collapse and is incorrect in the
strong interaction regime where higher order correlations start to
play an important role.

A realistic Hamiltonian in quasiparticle representation which accounts
for the mean-field, monopole pairing and the isoscalar,
multipole-multipole long-range part of the residual interaction has
the following form

\begin{equation}
  H = H_{qp} + H_{res}, 
\end{equation}

where

\begin{equation}
  H_{qp} = \sum\limits_{jm} \varepsilon_j \alpha_{jm}^\dagger  \alpha_{jm}
\end{equation}

\begin{widetext}
\begin{align}
  \label{eq:hamiltonian_general}
  & H_{res} = - \frac{1}{2}\sum_{\lambda \mu 12 1^\prime 2^\prime
    \rho\tau } \kappa ^\lambda_0 \frac{{(-1)^{\lambda  - \mu } }}{{\pi _\lambda ^2 }}
  f_{12}^{\lambda }( \tau) f_{1'2'}^{\lambda }(\rho\tau) \times \nonumber\\
  & \left\lbrace  \frac{1}{2}u_{12 }^{ + }(\tau)
    [ {A^{\tau  + } ( {12 ; \lambda \mu } ) +
      ( - 1)^{\lambda  - \mu } A^\tau  ( {12 ;\lambda  - \mu } )} ]
    + v_{12 }^{-}(\tau ) 
    B^\tau  ( {12 ; {\lambda \mu } })\right\rbrace \times \nonumber \\
  &\left\lbrace  \frac{1}{2}u_{1^\prime 2^\prime  }^{+ }(\rho\tau) [
    A^{\rho\tau  + } ( 1^\prime  2^\prime ; \lambda  - \mu ) + ( {
      - 1} )^{\lambda  - \mu } A^{\rho\tau } ( 1^\prime 2^\prime ;
    \lambda \mu )] +  v_{1^\prime 2^\prime }^{-}(\rho\tau) B^{\rho\tau}( {1
      ^\prime  2^\prime  ; \lambda  - \mu  } )\right\rbrace.
\end{align}

\end{widetext}

Expressions for the quantities in Eqs. \eqref{eq:normalization_factor}
and \eqref{eq:energy_expansion} using the above Hamiltonian, in the
case of one-nucleon species, are given in
App. \ref{sec:appendix1}. There it is shown that under the limiting
conditions listed in Sec. \ref{sec:lmg_basics} these expressions
coincide with the ones obtained within the LMG model with $2p-2h$
admixtures into the ground state. The numerical solution of
Eq. \eqref{eq:eq_general} is left out as a subject for a future study.

A common technique, which allows to mimic the dynamics govern by the
Hamiltonian \eqref{eq:hamiltonian}, from one side, and provides a
tractable way of evaluating these quantities, from the other, is to
have recourse to exactly solvable models which, in our case, would
ideally incorporate pairing and quadrupole terms. This however proves
impossible due to the fact that these two interactions are associated
with incompatible symmetry groups \cite{2003_rowe}. In this paper we
used the simplistic and widely used LMG model as a testbed for proving
the correctness of our idea.

\section{Solution within the Lipkin-Meshkov-Glick model}
\label{sec:lmg}

\subsection{LMG model basics}
\label{sec:lmg_basics}

In order to access the utility of different approaches and to prove
the usefulness of the proposed scheme we limit the configuration space
and simplify the inter-nucleon interaction to monopole-monopole one as
suggested by Lipkin, Meshkov and Glick \cite{1965_lipkin, 2010_maruhn}.
 This setting permits comparisons between the rates of
convergence of different approximation methods, including the hereby
described, to the exact solution.

In this model the interaction of $N$ particles on 2 quantum levels is presented by
the following Hamiltonian

\begin{equation}
  \label{eq:hamiltonian}
  H = H_0 + V; \;   H_0 = \varepsilon J_0; \;  V = \frac G2 (J_++J_{-})^2,
\end{equation}

where

\begin{align}
  \label{eq:lmg_operators}
  & J_+ =  \sum_i a_{1i}^\dagger a_{-1i}, \nonumber\\
  & J_- =  \sum_i a_{-1i}^\dagger a_{1i}, \nonumber\\
  & J_0 = \frac 12 \sum_i (a_{1i}^\dagger a_{1i} - a_{-1i}^\dagger a_{-1i} )
\end{align}

are analogous to the raising, lowering and angular momentum'
$z$-component of the quasi-spin algebra respectively, $a^{\dagger}$
represents the particle creation operator, the suffix $\pm 1$ denotes
the upper or lower level, $\varepsilon$ is the energy gap between the
two levels and $G$ is the interaction strength.

We shall also make use of the operators
\begin{equation}
  s^+_n  =  a^\dagger_{1n}a_{-1n};  s^-_n  =  a^\dagger_{-1n}a_{1n}; s^0_n = \frac 12 (a_{1n}^\dagger a_{1n} - a_{-1n}^\dagger a_{-1n}).
\end{equation}

The Hamiltonian \eqref{eq:hamiltonian} can be considered as a
specialization (up to a constant term) of the more general one
\eqref{eq:hamiltonian_general} under the following simplifications:

\begin{itemize}
\item pairing is switched off;
\item the number of levels is reduced to only two $\lbrace-1,1\rbrace$,
  each with a particle capacity of $N$;
\item the monopole-monopole part of the interaction is only considered;
\item the terms in the Hamiltonian \eqref{eq:hamiltonian_general},
  quadratic with respect to operator $B$ (Eq.\eqref{eq:B}), are
  neglected.
\end{itemize}

The operators \eqref{eq:lmg_operators} can be expressed by the ones
defined in Eqs.  \eqref{eq:A} and \eqref{eq:B} in the following way

\begin{equation}
  J_+ = -\sqrt{N}A^\dagger, \;\; J_- = -\sqrt{N}A,
\end{equation}

\begin{equation}
  J_0 = \frac 12 \sqrt{N} \left( B_{+1}+B_{-1} - \sqrt{N} \right),
\end{equation}

where 

\begin{equation}
  A^\dagger = A^\dagger((-1)(+1);00),
\end{equation}

\begin{equation}
  B_{+1} = B((+1)(+1);00),
\end{equation}

\begin{equation}
  B_{-1} = B((-1)(-1);00).
\end{equation}

The interaction strengths in Eqs. \eqref{eq:hamiltonian} and
\eqref{eq:hamiltonian_general} are related as

\begin{equation}
  G = - \frac{\kappa_0^0}{N}.
\end{equation}

\subsection{Exact solution}

The exact solution of the many-body problem is obtained as a
superposition of the (normalized) states $|n\rangle$ with $0,1,2 \ldots , N$
particles on the upper level ($|0\rangle \equiv \rangle$):

\begin{equation}
  \label{eq:exact_wf}
  | \Psi \rangle = \sum_{n}c_{n} | n \rangle.
\end{equation}

The weights $c_{n}$ are readily obtained by solving the eigenvalue
problem

\begin{equation}
  \sum_{n^\prime}\langle n | H | n^\prime \rangle c_{n^\prime} = Ec_{n}.
\end{equation}

The non-zero elements of the matrix on the left-hand side of the
above equation evaluate to

\begin{align}
  &\langle n|H|n\rangle = \\
  & = \left(-\frac N2 + n\right)\varepsilon + G\left( -\frac N2 + n + (n+1)(N-n) \right),
\end{align}

\begin{equation}
  \langle n|H|n+2\rangle = \frac G2 \sqrt{n(n-1)(N+2-n)(N+1-n)}.
\end{equation}

For the ground-state total energy we then obtain

\begin{equation}
  E = \sum_{n}(c_{n}^{(0)})^{2}\langle n| H | n\rangle + 2\sum_{n}c_{n}^{(0)}c_{n+2}^{(0)}\langle n| H |n+2\rangle.
\end{equation}

We shall further present the solutions for excited states, containing
only one particle on the upper level and one hole on the lower one,
i.e.

\begin{equation}
  \label{eq:wf_1p1h}
  |1p1h\rangle_{m} =  \left( \sum_i (\psi_i^{(m)} s^{+}_i - \varphi_i^{(m)} s^{-}_i) \right) \rangle.
\end{equation}

\subsection{RPA}
Employing the RPA, i.e.

\begin{equation}
  \label{eq:rpa}
  \langle| s_{i}^{-}, s_{i^{\prime}}^{+}  |\rangle = \delta_{ii^{\prime}}
\end{equation}

one obtains the well-known equation for the excitation energies and
the forward and backward amplitudes:

\begin{align}
\label{eq:2x2_rpa}
  &\left( {\begin{array}{*{20}c}
   {\delta_{ii^{\prime}}(\varepsilon - G) + G} & { -G(1- \delta_{ii^{\prime}}) }  \\
   {-G(1- \delta_{ii^{\prime}}) } & {\delta_{ii^{\prime}}(\varepsilon - G) + G}  \\
\end{array}} \right) \left( {\begin{array}{*{20}c}
   {\psi^{(m)}_{i^{\prime}} }  \\
   { \varphi^{(m)}_{i^{\prime}} }  \\
\end{array}} \right) = \nonumber\\ 
  &= \omega \left( {\begin{array}{*{20}c}
   {\delta_{ii^{\prime}}} & {0 }  \\
   {0 } & {-\delta_{ii^{\prime}}}  \\
\end{array}} \right)  \left( {\begin{array}{*{20}c}
   {\psi^{(m)}_{i^{\prime}} }  \\
   {\varphi^{(m)}_{i^{\prime}} }  \\
\end{array}} \right),
\end{align}

which together with the normalization of the wave functions
\eqref{eq:wf_1p1h} yields a collective solution 

\begin{equation}
  \psi = \frac{1}{\sqrt{N}}\frac{1+\chi/2 + \omega_0/\varepsilon} {\sqrt{\left(1 + \frac{\omega_0}{\varepsilon}\right) \left(1+\chi + \frac{\omega_0}{\varepsilon}\right)}},
\end{equation}

\begin{equation}
\varphi = \frac{1}{\sqrt{N}}\frac{\chi/2}{ \sqrt{ \left(1 + \frac{\omega_0}{\varepsilon}\right) \left(1+\chi + \frac{\omega_0}{\varepsilon}\right)} }, 
\end{equation}

where

\begin{equation}
  \chi = \frac{2G(N-1)}{\varepsilon}, \;\;  \omega_0 = \varepsilon \sqrt{1+\chi}.
\end{equation}

Since the phonon amplitudes are independent of the particle-hole pair
$i$ which they refer to due to the symmetry of the model and we are
interested in the collective solution only the wave function
\eqref{eq:wf_1p1h} can be rewritten in the more compact form which we
shall further use

\begin{equation}
  \label{eq:wf_1p1h_J}
  |1p1h\rangle =  \left(  \psi J_+ - \varphi J_- \right) \rangle.
\end{equation}

The particle occupation of the lower LMG level is easy obtained as

\begin{equation}
  \rho = 1 - \varphi^{2}.
\end{equation}

\subsection{ERPA}

The condition \eqref{eq:rpa} disregards some aspects of the nature of
the excited states \eqref{eq:wf_1p1h}, in particular the fact that the
number of particle-hole states in the ground state may be
non-negligible if sufficiently strong interaction is applied. In a
broader context, than the hereby considered, Hara \cite{1964_hara}
suggested to include explicitly the number of quasiparticles on each
level, which turned out to have a dramatic effect on the collective
properties of the low-lying states in open-shell even-even nuclei
\cite{1993_karadjov, 2010_mishev}. Adapting this approach to the LMG
model we can write

\begin{equation}
  \label{eq:erpa}
  \langle| n_{+1} |\rangle = N\rho,\; \langle| n_{-1} |\rangle = N(1- \rho).
\end{equation}

Eq. \eqref{eq:rpa} then transforms to

\begin{equation}
  \label{eq:erpa1}
  \langle| s_{n}^{-}, s_{n^{\prime}}^{+}  |\rangle = \delta_{nn^{\prime}}(1-2\rho).
\end{equation}

Analogous to Eq. \eqref{eq:2x2_rpa} in the current context is
the following one

\begin{equation}
\label{eq:2x2_erpa}
  \left( {\begin{array}{*{20}c}
   {A_{ii^{\prime}} } & {B_{ii^{\prime}} }  \\
   {B^\ast_{ii^{\prime}} } & {A^\ast_{ii^{\prime}}}  \\
\end{array}} \right) \left( {\begin{array}{*{20}c}
   {\psi^{(m)}_{i^{\prime}} }  \\
   { \varphi^{(m)}_{i^{\prime}} }  \\
\end{array}} \right) = 
  \omega \left( {\begin{array}{*{20}c}
   {U_{ii^{\prime}} } & {0 }  \\
   {0 } & {-U^\ast_{ii^{\prime}}}  \\
\end{array}} \right)  \left( {\begin{array}{*{20}c}
   {\psi^{(m)}_{i^{\prime}} }  \\
   { \varphi^{(m)}_{i^{\prime}} }  \\
\end{array}} \right),
\end{equation}

where

\begin{equation}
  A_{nn^{\prime}} = G(1-2\rho)^2 - \delta_{nn^{\prime}}(1-2\rho)(G-\varepsilon),
\end{equation}

\begin{equation}
  B_{nn^{\prime}} = G(1 - 2\rho)(\delta_{nn^{\prime}} +  2\rho - 1 ),
\end{equation}

and 
\begin{equation}
  U_{nn^{\prime}} = \delta_{nn^{\prime}}(1-2\rho).
\end{equation}

The solution of these equations is obtained to be

\begin{equation}
  \psi = \frac{1}{\sqrt{N(1-2\rho)}}\frac{1+\chi/2 + \omega_0/\varepsilon} {\sqrt{\left(1 + \frac{\omega_0}{\varepsilon}\right) \left(1+\chi + \frac{\omega_0}{\varepsilon}\right)}},
\end{equation}

\begin{equation}
  \label{eq:erpa_phi}
 \varphi = \frac{1}{\sqrt{N(1-2\rho)}}\frac{\chi/2}{ \sqrt{ \left(1 + \frac{\omega_0}{\varepsilon}\right) \left(1+\chi + \frac{\omega_0}{\varepsilon}\right)} },
\end{equation}

where 

\begin{equation}
  \label{eq:erpa_omega0}
  \omega_0^2 = \varepsilon^2(1+\chi),\;\;   \chi = \frac{2G}{\varepsilon}[(1-2\rho)N - 1].
\end{equation}

The system of equations closure is insured by the additional relation:

\begin{equation}
  \label{eq:erpa_rho}
  \rho = \frac{(\varphi)^{2}}{1+2 (\varphi)^{2} }.
\end{equation}

In order to obtain the correct ERPA solution one needs to solve the
system of coupled equations \eqref{eq:erpa_phi},
\eqref{eq:erpa_omega0}, and \eqref{eq:erpa_rho}.

\subsection{Phonon vacuum solution}
\label{sec:phonon_vacuum}

Finally, the featured method that we examine
(conf. Sec. \ref{sec:formalism}) translates in the language of the LMG
model in the following way. The wave function
\eqref{eq:def_transform_gs} assumes the form

\begin{equation}
  \label{eq:wf_lmg}
  |\rangle = N_{0} e^{\frac 12 S \sum_{ii^{\prime}}  s_{i}^+s_{i^{\prime}}^+} \rangle = N_{0} e^{\frac 12 S J_+^2} \rangle.
\end{equation}

Up to arbitrary order $n \leq N/2$ the ground-state energy is obtained to be

\begin{align}
  \label{eq:ph_vac_energy}
  & \langle| H |\rangle = N_{0}^{2}\sum_{n}\frac{1}{(n!)^{2}} \left(\frac S2 \right)^{2n} \langle J_{-}^{2n}HJ_{+}^{2n} \rangle + \\
  & 2N_{0}^{2}\sum_{n}\frac{n}{(n!)^{2}}\left( \frac S2 \right)^{2n-1}\langle J_{-}^{2n-2}HJ_{+}^{2n} \rangle \nonumber
\end{align}

with

\begin{align}
  N_{0}^{2} = \left[ \sum_{n} \frac{1}{(n!)^{2}} \left(\frac S2 \right)^{2n} \langle J_{-}^{2n}J_{+}^{2n} \rangle  \right]^{-1}
\end{align}

The variational equation $\partial_{S}\langle | H | \rangle = 0$ then
yields the following problem

\begin{widetext}
\begin{align}
  \label{eq:lmg_var_eq}
  & N_{0}^{2}\sum_{n}\frac{n}{(n!)^{2}} \left(\frac S2 \right)^{2n-2} \left[ \frac S2 \langle J_{-}^{2n}HJ_{+}^{2n} \rangle + (2n-1)\langle J_{-}^{2n-2}HJ_{+}^{2n} \rangle \right] +  \\
  & \partial_{S}N_{0}^{2}\sum_{n}\frac{1}{(n!)^{2}} \left(\frac S2 \right)^{2n-1} \left[ \frac S2 \langle J_{-}^{2n}HJ_{+}^{2n} \rangle + 2n\langle J_{-}^{2n-2}HJ_{+}^{2n} \rangle \right] = 0. \nonumber
\end{align}
\end{widetext}

Respectively, the energy of the $1p-1h$ excited state is evaluated as

\begin{align}
  \label{eq:lmg_omega}
  \omega =  \psi^{2} \langle | J_{-} H J_{+} | \rangle - 2 \psi\varphi \langle | J_{-} H J_{-} | \rangle + \varphi^{2} \langle | J_{+} H J_{-} | \rangle.
\end{align}

The expressions for the Hamiltonian average values in
Eqs.\eqref{eq:lmg_var_eq} and \eqref{eq:lmg_omega} are given in
App. \ref{sec:appendix2}. The forward and backward phonon amplitudes
in Eq.\eqref{eq:lmg_omega} are obtained by applying the normalization
condition for the one-phonon state

\begin{widetext}
\begin{align}
  \label{eq:lmg_norm_cond_general}
  & N_0^2 \left( \psi^2-\varphi^2\right)\sum_{n}\left( \frac{1}{n!} \right)^2 \left( \frac S2 \right)^{2n}\left(N - 4n\right)\left\langle J_-^{2n}J_+^{2n} \right\rangle =1,
\end{align}
\end{widetext}

along with the definitive equation

\begin{equation}
  Q |\rangle = 0,
\end{equation}

resulting in the relation

\begin{equation}
  \label{eq:lmg_psi_phi_s}
  \left[ \varphi - (N-1)\psi S \right]^2 + 6\psi^2 S^{4}(N-1)(N-2)= 0.
\end{equation}

Note that the latter relation is independent of the expansion order
$n$.

If we truncate the exponent expansion \eqref{eq:wf_lmg} to first
order, i.e. allow for $2p-2h$ admixtures only in the ground-state wave
function, we obtain

\begin{equation}
  \label{eq:norm_factor_lmg}
  N^{2}_{0} = \frac{1}{1+ \frac 12 N(N-1)S^2 }.
\end{equation}

The variational problem then is rewritten as

\begin{equation}
  \label{eq:variation_lmg}
    \delta \left( N_{0}^2 \left\langle (1 + \frac 12 SJ_-^2 ) H (1 + \frac 12 SJ_+^2  )  \right\rangle \right) = 0.
\end{equation}

The expressions for the relevant quantities in this equation are given
in App. \ref{sec:appendix1}. The structure coefficient $S$ in
Eq.\eqref{eq:wf_lmg} is related to those in
Eq.\eqref{eq:oper_nqp} in the following way:

\begin{equation}
  S = -\frac 4N S_0(+1-1;+1-1).
\end{equation}

Performing the variation \eqref{eq:variation_lmg} we get the following
simple quadratic equation for $S$:

\begin{equation}
  1 + \left(2 \frac{\varepsilon}{G} + 2N - 4 \right)S  - \frac 12 (N^2-N)S^{2} =0.
\end{equation}

\begin{figure*}[t!]
  \includegraphics[scale=0.425]{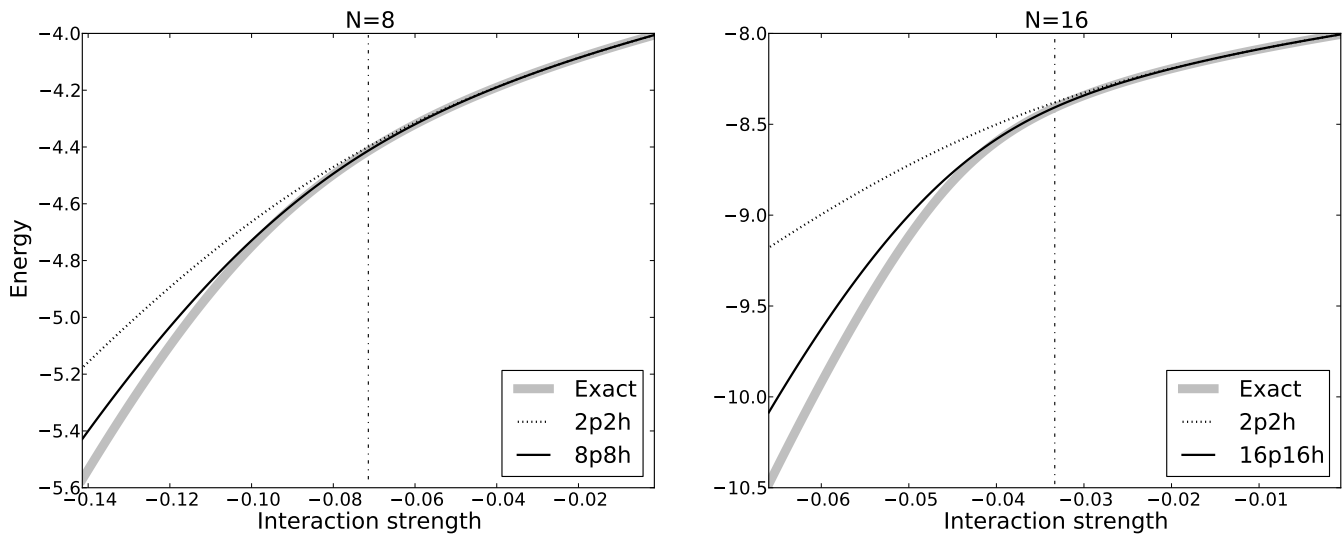}
  \caption{ \label{fig:e}Ground-state energy (MeV) in a two-level LMG
    model systems with $N=8$ (left panel) and $N=16$ (right panel)
    particles as function of the interaction strength
    $G$(MeV). Solutions obtained using the phonon vacuum explicit
    variation with $2p-2h$ and up to $16p-16h$ admixtures into the
    ground-state wave function are compared to the exact one. The
    vertical dash-dotted line indicates the strength at which the RPA
    experiences a collapse. The energy gap $\varepsilon$ between the
    two LMG levels is set to 1 MeV.}
\end{figure*}

The ground-state energy in this case evaluates to

\begin{widetext}
\begin{equation}
  \label{eq:energy_ph_vac}
  E =  N_{0}^2 \left[ \langle H \rangle + N(N-1)GS +  \frac 14 (\varepsilon - G) ( - N + 4) N(N-1) S^2 + GN(N-1)^{2}S^{2} \right].
\end{equation}
\end{widetext}

\begin{figure}[b]
  \includegraphics[scale=0.45]{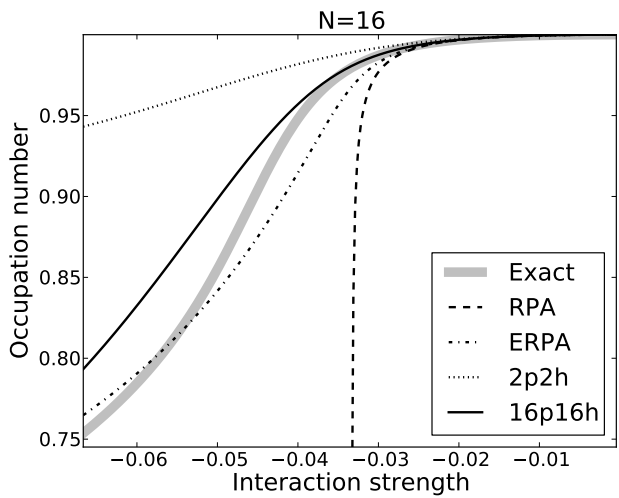}
  \caption{ \label{fig:rho}Same as Fig. \ref{fig:e} but for the
    occupation number on the lower LMG model level for a system of $N=16$ particles. RPA and ERPA
    curves are also plotted.}
\end{figure}

The transition between the quantities obtained using the realistic
Hamiltonian\eqref{eq:hamiltonian_general} and the LMG ones with
$2p-2h$ correlations only is performed in App.\ref{sec:appendix1}.

\section{Numerical results}
\label{sec:numerics}

\begin{figure*}
\includegraphics[scale=0.45]{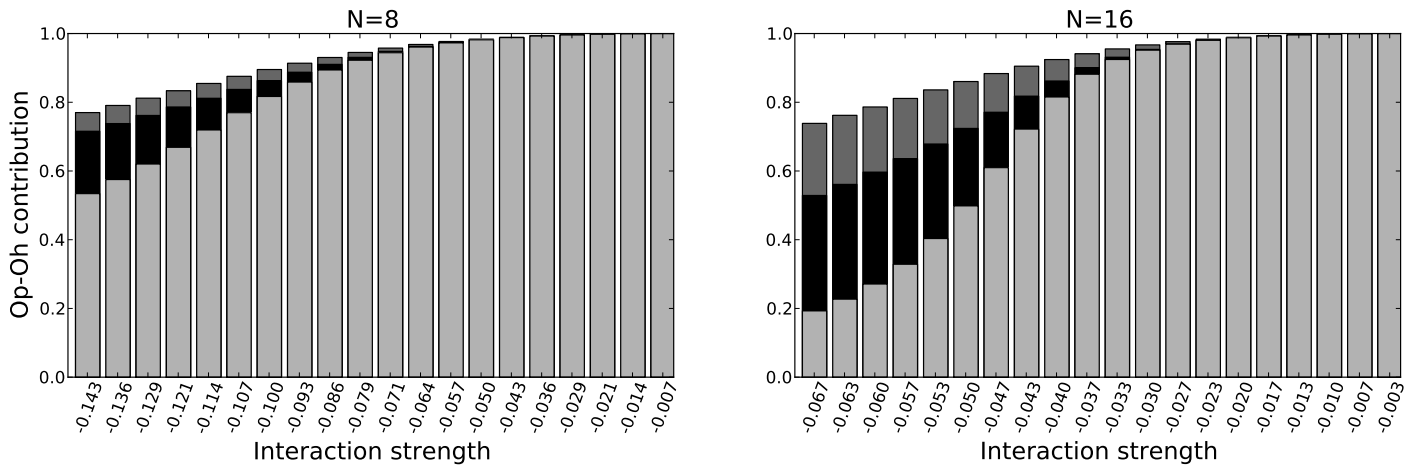}
\caption{\label{fig:components} The $0p-0h$ contribution to the exact
  ground-state wave function from Eq.\eqref{eq:exact_wf} (light gray),
  the phonon vacuum solution from Eq.\eqref{eq:wf_lmg} with $2p-2h$
  admixtures (dark gray) and with correlations of order $N/2$ (black)
  as function of the interaction strength $G$ for LMG model systems
  with $N=8$ (left panel) and $N=16$ (right panel) particles.}
\end{figure*}

The three approximations, presented in the previous section, are
compared with the exact solution based on the ground and first excited
state energies (Fig.\ref{fig:e}) as well as based on the occupation
particle density (Fig.\ref{fig:rho}) on the lower LMG model level in
the ground state.  We assigned $N\leq16$ particles to the system for a
simple reason - if we consider the two levels of the LMG model
representing the valence sub-shells in the nucleus then these would be
the source of the major effects for the low-lying states and $N=16$
would mimic the neutron or proton subsystem of a nucleus from the
middle of $sdgh$ major shell.

The first observation in Fig. \ref{fig:e} is the clearly designated
critical RPA strength

\begin{equation}
  \label{eq:g_rpa_crit}
  G_{crit} = -\frac{\varepsilon}{2(N-1)},
\end{equation}

which separates the region where a real RPA solution can be found from
the one, in which only a complex solution is obtained. It is also
worthwhile to notice that the RPA point of collapse stands at the
onset of the transition between the two nearly linear sections of the
exact solution for the ground-state energy, which are more
distinguished in systems with a larger number of particles.

\begin{figure*}
  \includegraphics[scale=0.425]{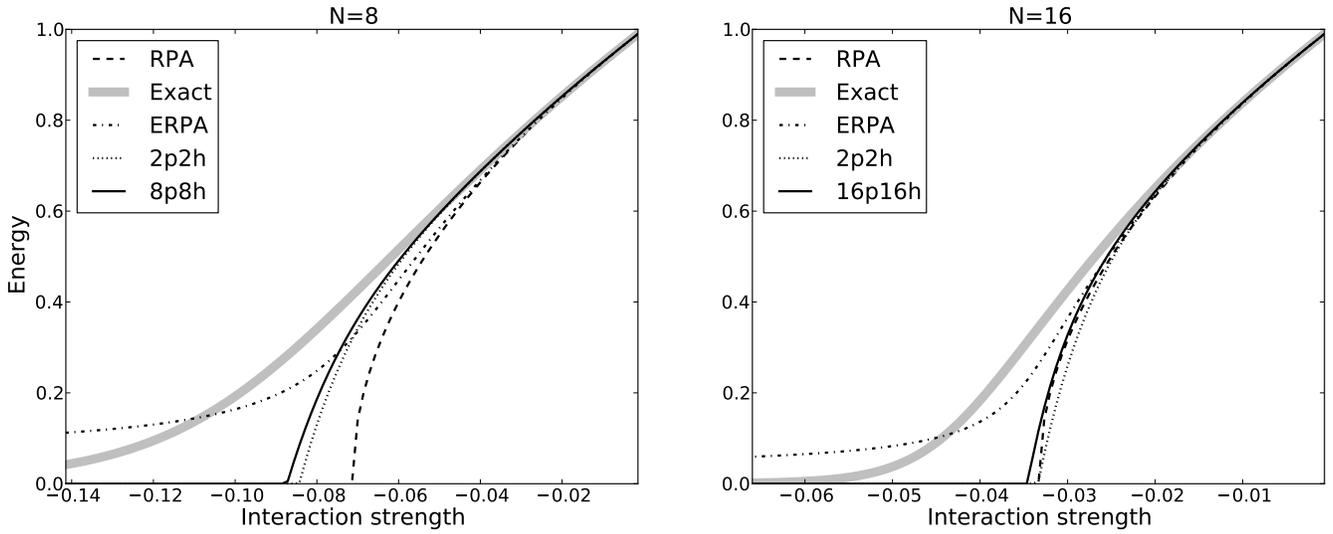}
  \caption{ \label{fig:e_excited} Same as Fig. \ref{fig:e} but for the
    energy of the first excited state. RPA and ERPA curves are also
    plotted.}
\end{figure*}

On the other hand, the explicit variation of the phonon vacuum with
$2p-2h$ admixtures only in the ground state yields solution at any
$G$. However, as seen from Figs. \ref{fig:e} and \ref{fig:rho},
increasing the interaction strength beyond the RPA critical point
causes progressive divergence of this solution from the exact one for
both the ground-state energy and the occupation number. This
divergence exacerbates incrementing the number of particles in the
system. Adding higher order terms to the energy functional in
Eq.\eqref{eq:ph_vac_energy}, which account for further correlations
effects, greatly improves the results bringing the energy of the
phonon vacuum and the occupation number closer to the exact value. The
importance of the multi-particle-hole admixtures to the ground state
in the strong interaction regime is illustrated in
Fig. \ref{fig:components} where the weight of the ground state's
$0p-0h$ component is plotted as function of the interaction strength
$G$. Dispite the reasonable agreement for the ground-state energies in
this regime the wave function of the phonon vacuum yields a
substantially less correlated state than the true ground state.The
prescription of the variational principle for monotonic convergence is
easily seen in Figs. \ref{fig:e},\ref{fig:rho}, \ref{fig:components}
and \ref{fig:e_excited} as the higher order correlations bring the
result closer to the exact solution but never overbinding the
ground state.

The behaviour of the first excited state's energy within the phonon
vacuum variational approach exhibits a collapse at interaction
strengths far weaker than those at which the first excited state
energy starts to diminish (see Fig.\ref{fig:e_excited}). The position
of this point is very much independent of the order of
multi-particle-hole correlations included in the phonon vacuum state
in cases when this point is near $G_{crit}$
(Eq. \eqref{eq:g_rpa_crit}) as in the weak interaction regime
admixtures beyond the $2p-2h$ ones contribute relatively little (see
Fig. \ref{fig:components}). The obvious reason for this collapse, as
noted in \cite{1999_hirsch}, can be attributed to the higher order
particle-hole admixtures contributing in the structure of the first
excited state which start to be an important factor as the interaction
becomes stronger.

As opposed to the RPA, a real ERPA solution, is found everywhere in
the range of $G$ values considered. In the interval $(G_{crit},0]$ it
  performs just as well as the RPA does except for strengths close to
  $G_{crit}$. In the strong interaction regime the ERPA gives rather
  good results both for the particle occupation and the first excited
  state energy (see Figs. \ref{fig:rho} and \ref{fig:e_excited}). Near
  and beyond $G_{crit}$ it predicts higher depletion of the lower LMG
  level (see Fig.\ref{fig:rho}) and correspondingly, as seen in
  Fig. \ref{fig:e_excited}, it gives lower value for the energy of the
  first excited state. At interactions twice as strong as the RPA
  point of collapse the first excited state's energy is found to be a
  bit higher compared to the exact solution. This result is to be
  expected given that the particle occupation number at such strengths
  within the ERPA is overestimated. It is worth noticing that in the
  strong interaction regime a solution of the system of coupled
  equations \eqref{eq:erpa_phi}, \eqref{eq:erpa_omega0} and
  \eqref{eq:erpa_rho} is found for an occupation number whose value
  $\rho$ is in very close proximity to the critical occupation number

\begin{equation}
  \label{eq:rho_erpa_crit}
  \rho_{crit} = \frac 12 - \frac {1}{2N} (1-\frac{\varepsilon}{2G}).
\end{equation}

\section{Conclusion and outlook}
\label{sec:conclusion}

In this work we initiated the development of a variational approach
for approximating the ground state of nuclei using a wave function in
the form of the phonon vacuum state, tailored to take into account the
action of the long-range residual forces. Essentially it is an attempt
to provide a controlled succession of approximations for estimating
the contributions from different multi-particle-hole
admixtures. Applying our idea to the very schematic LMG model, we
showed the superiority of our approach over the RPA, based on
comparison with the exact solution for the ground and first-excited
state energies as well as based on the particle occupation
numbers. Alongside that we adapted the ERPA \cite{1964_hara} to the
LMG model to show its high utility within a wide range of interaction
strengths. As far as the LMG model is able to simulate real nuclei, we
conclude the following:
  \begin{itemize}
  \item The RPA point of collapse separates two types of system's
    behavior - the weak and strong interaction regimes.
  \item The RPA provides an accurate and computationally efficient
    method to treat nuclei until the point of collapse inherent to
    this approximation. 
  \item The ERPA improves over the standard RPA as it yields solution
    for any interaction strength. The calculated value for the first
    excited state's energy is the closest to the exact one amongst all
    considered approximations. The ground state occupation numbers
    within this approximation are also reproduced with reasonable
    accuracy.
  \item Explicitly varying the phonon vacuum state one obtains a very
    good accuracy to the ground-state energy as the result improves
    monotonically adding higher order correlations into the ground
    state. In the region beyond the RPA point of collapse, multiple
    particle-hole admixtures higher than $2p-2h$ ones start to play
    very important role and at such strengths, as seen from
    Fig. \ref{fig:components}, the phonon vacuum solution
    significantly underestimates the degree of correlation. This
    effect is amplified with the increase of the number of particles
    in the system.
  \item The energy of the first excited state can be treated as an
    $1p-1h$ excitation over the ground state only in the weak
    interaction regime until the RPA point of collapse. Beyond that
    point it vanishes rapidly especially for systems with a larger
    number of particles and its proper description requires
    consideration within and extended configuration space for the
    excited states.
  \end{itemize}

Applying the presented approach to real nuclei one can expect to
obtain improved results than previously published for isotopes from
the transitional regions.Being a many-body wave function this approach
may be a good choice for describing phenomena involving many-particle
correlations as, for example, cluster configurations.  The influence
of the long-range residual forces on the mean field and the pairing
correlations in real nuclei is another perspective which the presented
development makes possible to realize and it is currently progressing.

\section*{Acknowledgments}
\label{sec:acknoledgements}

The author is indebted to Prof. V.V. Voronov for helpful comments and
valuable discussions during the course of this work. Thanks are also
due to Prof. R.V. Jolos for his useful remarks.

This work is partially supported by JINR grant No. 13-302-07.

\appendix

\section{}
\label{sec:appendix1}

Here we give expressions for the quantities in
Eq. \eqref{eq:energy_expansion} for single species of nucleons. A
transition between these expressions and the ones obtained within the
LMG model with $2p-2h$ correlations into the ground-state wave
function is performed.

\begin{equation}
  \label{eq:energy_expansion_first}
  \langle H \rangle =  \sum_{12\lambda} \kappa_{0}^{\lambda} (h_{12}^{\lambda})^{2},
\end{equation}

\begin{widetext}

\begin{equation}
  \label{eq:energy_expansion_second}
  \langle S^{\dagger}S \rangle = 2\sum_{1234 \lambda}\pi_{\lambda}^{2}S^{2}_{\lambda}(1234)  + 4\sum_{1234 \lambda \lambda_1} \pi_{\lambda_1}^{2}\pi_{\lambda}^{2} S_\lambda(1234)S_{\lambda_1}(2431)   \left\{ {\begin{array}{*{20}c}
        {3 } & {4 } & {\lambda }  \\
        {2 } & {1 } & {\lambda_1 }  \\
      \end{array}} \right\} ,
\end{equation}

\begin{equation}
  \label{eq:energy_expansion_third}
  \langle HS \rangle = 2\sum_{1234 \lambda}\pi_{\lambda}^{2}\kappa_0^\lambda H_{\lambda}(1234)S_{\lambda}(1234)  + 4\sum_{1234 \lambda \lambda_1} \pi_{\lambda_1}^{2}\pi_{\lambda}^{2} \kappa_0^\lambda H_\lambda(1234)S_{\lambda_1}(2431)   \left\{ {\begin{array}{*{20}c}
        {3 } & {4} & {\lambda }  \\
        {2 } & {1} & {\lambda_1 }  \\
      \end{array}} \right\},
\end{equation}

\begin{equation}
  \label{eq:SHS}
  \langle S^{\dagger}HS \rangle = \langle S^{\dagger}H_{1}S \rangle + \langle S^{\dagger}H_{2}S \rangle
\end{equation}

with

\begin{equation}
  \label{eq:SH1S}
  \left\langle S^{\dagger} H_1 S \right\rangle = 2\sum_{1234 \lambda}\pi_{\lambda}^{2}\varepsilon_{1234}S^{2}_{\lambda}(1234)  + 4\sum_{1234 \lambda \lambda_1} \pi_{\lambda_1}^{2}\pi_{\lambda}^{2} \varepsilon_{1234} S_\lambda(1234)S_{\lambda_1}(2431)   \left\{ {\begin{array}{*{20}c}
        {3 } & {4} & {\lambda }  \\
        {2 } & {1} & {\lambda_1 }  \\
      \end{array}} \right\} 
\end{equation}

\begin{align}
  &\langle S^{\dagger}H_{2} S \rangle= -8 \sum_{1234\lambda} \pi_{\lambda}^2 \kappa_0^{\lambda}  F_{\lambda}(1234) S_{\lambda_1}(1234) - 16 \sum_{1234\lambda \lambda_1} \pi_{\lambda}^2 \pi_{\lambda_1}^2 \kappa_0^\lambda F_{\lambda}(1234) S_{\lambda_1}(2431)\left\{ {\begin{array}{*{20}c}
        { 3} & {4 } & {\lambda }  \\
        {2 } & {1 } & {\lambda_{1} }  \\
      \end{array}} \right\} \\ \nonumber
  & + 16 \sum_{1234\lambda} \pi_{\lambda}^{2}R_{\lambda}(1234)S_{\lambda}(1234) + 32\sum_{1234\lambda\lambda_{1}}\pi_{\lambda}^{2}\pi_{\lambda_{1}}^{2}R_{\lambda}(1234)S_{\lambda_{1}}(2431)\left\{ {\begin{array}{*{20}c}
        { 3} & {4 } & {\lambda }  \\
        {2 } & {1 } & {\lambda_{1} }  \\
      \end{array}} \right\}\\\nonumber
  &- \left(\sum_{aa'J}(h_{aa'}^{J})^{2}\kappa_{0}^{J}\right)\left[2\sum_{1234\lambda}\pi_{\lambda}^{2}S_{\lambda}^{2}(1234) + 4\sum_{1234\lambda\lambda_{1}}\pi_{\lambda}^{2}\pi_{\lambda_{1}}^{2}S_{\lambda}(1234)S_{\lambda_{1}}(2431)\left\{ {\begin{array}{*{20}c}
        { 3} & {4 } & {\lambda }  \\
        {2 } & {1 } & {\lambda_{1} }  \\
      \end{array}} \right\} \right],
\end{align}

where
\begin{align}
  F_{\lambda}(1234) = \frac{h^\lambda_{34}}{\pi_\lambda^2}  \sum_{56} h^\lambda_{56}    \left[ S_\lambda(1256) + 2\sum_J \pi^{2}_{J}  \left\{ {\begin{array}{*{20}c}
        {5} & {6} & {\lambda }  \\
        {1 } & {2 } & {J }  \\
      \end{array}} \right\} S_J(5216) \right] 
\end{align}

and 
\begin{equation}
  R_{\lambda}(1234)=\left(\sum_{5J}\kappa_{0}^{J}(h_{15}^{J})^{2}\right) \frac{S_{\lambda}(1234)}{\pi_{33}}.
\end{equation}

\end{widetext}

Here we used the shortcut notation

\begin{equation}
  H_{\lambda}(1234) = h^{\lambda}_{12}h^{\lambda}_{34},
\end{equation}

where

\begin{equation}
  h_{12}^{\lambda} = \frac{f_{12}^{\lambda}u_{12}^{+}}{2}.
\end{equation}

In Eq.\eqref{eq:SH1S} $\varepsilon_{1234}$ stands for the energy of a
four-quasiparticle state, i.e.
$\varepsilon_{1234} = \varepsilon_{1}+\varepsilon_{2}+\varepsilon_{3}+\varepsilon_{4}$. In
Eq.\eqref{eq:SHS} the term in the Hamiltonian
\eqref{eq:hamiltonian_general}, quadratic with respect to the operator
$B$, defined in Eq.\eqref{eq:B}, is omitted.

Applying the considerations from Sec.\ref{sec:lmg_basics} and using
the relation

\begin{equation}
\left\{ {\begin{array}{*{20}c}
        {j } & {j } & {0 }  \\
        {j } & {j } & {0 }  \\
      \end{array}} \right\} = -\frac{1}{2j+1} = -\frac{1}{N},
\end{equation}

one obtains the corresponding LMG model expressions with $2p-2h$
correlations in the ground-state wave function:

\begin{equation}
  \langle H \rangle = (G - \varepsilon) \frac N2,
\end{equation}

\begin{equation}
  \langle S^{\dagger} S \rangle  = \frac 12 N(N-1) S^{2},
\end{equation}

\begin{equation}
  \langle H S \rangle  =  \frac 12 G N(N-1) S,
\end{equation}

\begin{equation}
  \langle S^\dagger H S \rangle = \left[ \frac{(\varepsilon-G)(4-N)N (N-1)}{4} + GN(N-1)^{2}\right]S^{2}.
\end{equation}

\section{}
\label{sec:appendix2}

In this Appendix we give the expressions for the Hamiltonian
  average values in the uncorrelated $\rangle$ and correlated
  $|\rangle$ ground states needed to solve Eqs. \eqref{eq:lmg_var_eq},
  \eqref{eq:lmg_norm_cond_general} and \eqref{eq:lmg_psi_phi_s} from
  Sec. \ref{sec:phonon_vacuum}.

\begin{equation}
  \langle J_{-}^{n}J_{+}^{n} \rangle = \frac{n!N!}{(N-n)!}
\end{equation}

\begin{equation}
  \langle J_-^n H J_+^{n+2}\rangle = \frac G2 \frac{(n+2)!N!}{(N-(n+2))!}
\end{equation}

\begin{equation}
  \langle J_-^n H J_+^{n}\rangle = \left[ (\varepsilon + G)(n - \frac N2) + G(n+1)(N-n) \right] \frac{n!N!}{(N-n)!}
\end{equation}

\begin{widetext}
\begin{equation}
  \langle | J_{-} H J_{+} | \rangle = N_{0}^{2} \sum_{n}\frac{1}{(n!)^{2}} \left(\frac S2 \right)^{2n} \langle J_{-}^{2n+1}HJ_{+}^{2n+1} \rangle + 2N_{0}^{2}\sum_{n}\frac{n}{(n!)^{2}} \left(\frac S2 \right)^{2n-1} \langle J_{-}^{2n-1}HJ_{+}^{2n+1} \rangle
\end{equation}

\begin{equation}
  \langle | J_{-} H J_{-} | \rangle = N_{0}^{2} \sum_{n}\frac{1}{(n!)^{2}} 2n(N-2n+1) \left(\frac S2 \right)^{2n} \langle J_{-}^{2n+1}HJ_{+}^{2n-1} \rangle + N_{0}^{2}\sum_{n}\frac{n}{(n!)^{2}} \left(\frac S2 \right)^{2n-1} 2n(N-2n+1) \langle J_{-}^{2n-1}HJ_{+}^{2n-1} \rangle
\end{equation}

\begin{align}
  & \langle | J_{+} H J_{-} | \rangle = N_{0}^{2} \sum_{n}\frac{1}{(n!)^{2}} 4n^{2}(N-2n+1)^{2} \left(\frac S2 \right)^{2n} \langle J_{-}^{2n-1}HJ_{+}^{2n-1} \rangle + \\
  & N_{0}^{2}\sum_{n}\frac{n}{(n!)^{2}} \left(\frac S2 \right)^{2n-1} (2n-2)2n(N-2n+3)(N-2n+1) \langle J_{-}^{2n-3}HJ_{+}^{2n-1} \rangle
\end{align}
\end{widetext}

\end{document}